\documentclass[journal, 11pt,onecolumn]{IEEEtran}

 \UseRawInputEncoding

\usepackage{times}

\usepackage{caption}  
\usepackage{algorithm}
\usepackage[noend]{algpseudocode}

\usepackage{multirow}

\ifCLASSINFOpdf
   \usepackage[pdftex]{graphicx}
  \else
     \usepackage[dvips]{graphicx}

 \usepackage{amssymb,amsmath}
 
\fi   
\usepackage{enumerate}
\usepackage{graphicx,wrapfig,lipsum}
\usepackage{amsmath}
\usepackage{comment}

\usepackage[caption=false,font=footnotesize]{subfig}

\usepackage{cite}

\begin{document}

\title{Micro-location for Smart Buildings  in the Era of Internet of Things }

\author{\IEEEauthorblockN{ Petros Spachos, Ioannis Papapanagiotou and Konstantinos Plataniotis}

\thanks{Petros Spachos is with the School of Engineering at the University of Guelph, Ontario, Canada, petros@uoguelph.ca.}
\thanks{Ioannis Papapanagiotou is with Platform Engineering at Netflix, Los Gatos CA, USA, ipapapa@ncsu.edu }
\thanks{Konstantinos N. Plataniotis is with the Department of Electrical and Computer Engineering at the University of Toronto, Ontario, Canada, kostas@ece.utoronto.ca} 
}
    \maketitle 
    \begin{abstract}
 
Micro-location plays a key role in the transformation of traditional buildings into smart infrastructure. Micro-location is the process of locating any entity with a very high accuracy, possibly in centimeters. Such technologies require high detection accuracy, energy efficiency, wide reception range, low cost, and availability. In this paper, we provide insights into various micro-location enabling technologies, techniques, and services, and discuss how they can accelerate the incorporation of the Internet of Things (IoT) in smart buildings. We cover the challenges and examine some signal processing filtering techniques such that micro-location enabling technologies and services can be thoroughly integrated with an IoT equipped smart building. An experiment with Bluetooth Low Energy (BLE) beacons used for micro-location is also presented.   

 \end{abstract}

\IEEEpeerreviewmaketitle

\section{Introduction.}
The interconnectedness of all things is continuously expanding. The aim is to have every individual interconnected with their surroundings, whether it be at home, at work, or in public spaces. Some of these services might include but are not limited to indoor mapping, personalized environment changes, such as lighting and temperature settings, as well as directed advertisement. For these systems to perform, it is essential to have reliable hardware and accurate data. Outdoor localization technologies, such as the Global Positioning System (GPS), do not work indoors due to the physical barriers that block the signal while it does not provide location data accurate enough for micro-location. Current solutions utilize Received Signal Strength Indicators (RSSI) to determine position. A variety of solutions that utilize RSSI have been proposed to provide location services for indoor environments, though each solution presents its own drawbacks.  Multiple technologies and  techniques  have been adapted to provide indoor location information, all of which attempt to overcome the noise and dynamics of a changing indoor environment.

A promising approach includes the effective use of the plethora of  Internet of Things (IoT) devices that are available in the market. Bluetooth Low Energy (BLE) beacons, usually referred to as beacons, are a promising candidate to improve indoor localization accuracy. They are small Bluetooth transmitters designed to attract attention to a specific location. As in many IoT based networks,  the performance of such networks relies on the network lifespan and accuracy. BLE beacons are a cheap, simple, and very scalable means of implementing indoor localization services. In recent years, BLE technology has grown in popularity and much more research has been developed in using it for indoor localization \cite{sadowski,smartBuildings,SikeMicro}. The fundamental operation of these beacons for localization purposes is based on RSSI techniques, where the received RSSI value is translated into a distance by using a best curve fit signal propagation model. BLE beacon protocols, such as iBeacon \cite{ibeaconpacket}, and Eddystone \cite{eddystonepacket}, provide the necessary information and configuration capabilities for micro-location. Along with the low power consumption of BLE, beacon devices are easily deployed and require low maintenance, hence their scalability for any complex indoor environment.

\begin{figure}[t!]
\centering
\includegraphics[width=5.5cm]{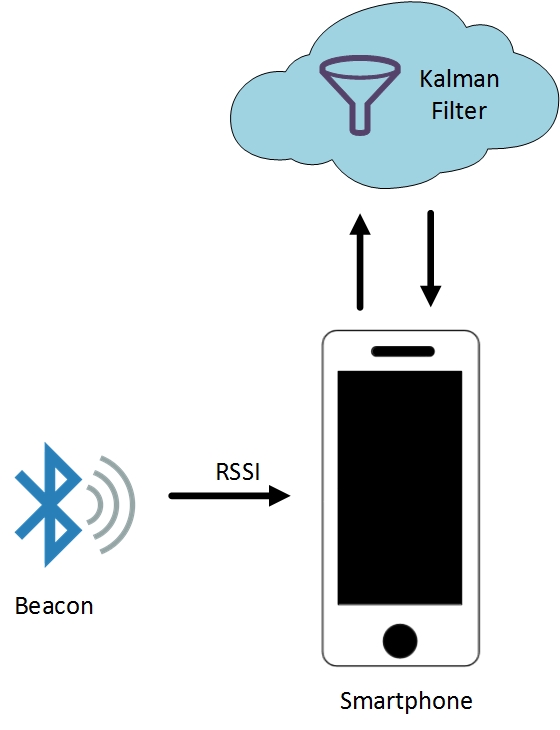}
\caption{Micro-location system using Kalman Filter.}\label{system}

\end{figure}

Intrinsic to any wireless technology, BLE beacons are highly susceptible to noise and interference. To overcome the effects of noise and dynamic changes to the physical environment, many methods devised around advanced positioning algorithms and filtering techniques have been adapted to beacon-based systems to improve the accuracy obtained in using RSSI localization techniques, as shown in Fig. \ref{system}. Some of the most common filter implementations are the Kalman filters as detailed in \cite{bordoy}. Kalman filtering has also been examined in the context of indoor localization \cite{zafari2017ibeacon}.  These filters provide a reasonably accurate state estimation and can be adjusted for changes is environmental/ process noise. Other filters such as the particle filter are used. Particle filters are highly accurate, but at the cost of greater computational complexity, hence the need for a client-server based model, as outlined in \cite{smartBuildings,particleFilt}. Positioning algorithms can also have an effect on beacon accuracy. The work presented in \cite{NN_knB} implements the K-Nearest Neighbor algorithm to calculate the position of the user. The experiments showed an average error of 1 m. Other algorithms, such as the Pedestrian Dead Reckoning (PDR) approach have been implemented with BLE beacons \cite{PDR}. In these experiments, the integration of smartphone sensors for data regarding step detection, step direction, and walking length, are combined with beacon calibration zones to provide a more accurate position. All techniques may provide different accuracy results and may behave differently depending on the environment, so it is important to note the characteristics of each tested environment when deciding on what technique to implement.
 
In this paper, we survey available wireless technologies for micro-location systems in a smart building. Then we discuss signal processing techniques and characteristics that can be used to improve micro-location performance, along with filtering approaches. We focus on the use of BLE beacons and through an experiment we discuss how they can enhance micro-location.
 
\section{Smart Buildings with Internet of Things Technologies.}

The IoT revolution has brought a swarm of continuously interconnected and sensor-packed devices opening a vast number of opportunities in equipping existing infrastructures. IoT has enabled applications that transform facilities to intelligent spaces able to critically affect and improve productivity and life quality of the occupants. Reducing energy costs, detecting and building knowledge based on human patterns as well as improving the human-building interaction is only some cases in point.

The Institute for Building Efficiency \cite{IBE} defines smart buildings as the buildings that can provide low-cost services such as air conditioning, heating, ventilation, illumination, security, sanitation and various other services to the tenants without adversely affecting the environment. This requires the collaboration of multiple sensors that form a building's IoT ecosystem. The basic motive behind the construction of smart buildings is to provide the highest level of comfort and efficiency. At the same time, the interconnection of the automation systems can assist with the disaster management and provide emergency services. For example, the collaboration of the fire system with the air conditioning system can create an environment where a fire will not expand in the rest of the building.

To the end, indoor-focused Location-Based Services (LBS) is the fundamental component for providing a tenant to building interaction. LBS provide the ability to efficiently track occupants in real-time. They either attempt to estimate the users' 2D coordinates, which is referred to as micro-location or assign the user in the locality of certain points of interest (PoI), which is known as proximity sensing.

The integration of smart buildings with IoT creates a number of challenges. A smart building with an IoT ecosystem requires mainly three components: the sensors, the integration, and the actuators. The sensors must be connected to a reliable highly available network that optimally can self-diagnose and heal. The integration is probably the part where innovation is nowadays taking place. The integration consists of some software that would receive the input from the sensors, process and analyze, and provide some actuator as a service to the tenants.  For example, unlocking a door, switching on the TV, calling the elevator or configuring the room temperature based on the needs.

\section{Overview of micro-location Systems.}
This section describes the wireless technologies and the radio signal features that can be used for micro-location in a complex indoor environment, followed by a brief description of the techniques that can be used to extract information location from the wireless signals.
 
\subsection{Wireless Technologies.}
Micro-location systems can leverage existing wireless infrastructure for micro-location to minimize the cost or may require a specific wireless deployment~\cite{sadowski}. By wireless technologies, we refer both to high-frequency technologies as well as low frequency. For example, the most common high-frequency wireless technologies that have been used in a micro-location deployment are WiFi \cite{wifi}, Zigbee \cite{zigbee}, RFID \cite{rfid} and Bluetooth \cite{blue5}. However, low-frequency technologies like the ones based on physical light have also seen some research and commercial use \cite{kusens2017electronic}. For example, Light Fidelity, or else Li-Fi is one of the wireless technologies in the form of Visible Light Communication (VLC) technology. These technologies have been used successfully in the past for indoor location and navigation and their popularity among the IoT devices make them an ideal solution for micro-location as well. There are also technologies such as WiFi HaLow \cite{halow}, BLE version 5.0 \cite{blue5} and LoRaWAN \cite{lorawan} which are specifically designed for IoT devices.
 
\begin{itemize}

\item \textbf{IEEE 802. 11 - WiFi.} The IEEE 802.11 standard \cite{wifi}, commonly known as WiFi, is among the most popular technologies used for localization when the GPS is inadequate. The great distribution of access points and signal availability at indoor environment make it easy to collect the received signals from various access points and calculate the location of the receiver. The indoor transmission range can vary from 3.3  m with a bandwidth of 6.7 Gbit/s (IEEE 802.11 ad), up to 70 m with a bandwidth of 600 Mbit/s (IEEE 802.11 n), while it can operate in 2.4 GHz, 5 GHz, and 60 GHz.
 
WiFi networks are deployed for communication, hence data rate and connectivity are important while localization is not their priority. Also, WiFi networks are designed for a plethora of devices from smartphone and laptops to phablets and smartwatches. This is a trade-off for micro-location techniques. The availability of the WiFi signals and WiFi-enabled devices is an advantage for micro-location as the number of the portable devices and potential reference points for localization increases. Advanced signal processing techniques can be used to improve the quality of the WiFi signals for localization. At the same time, there is no need for extra hardware deployment with WiFi technology.

On the other hand, IoT devices have unique characteristics, such as the size and the limited energy resources, that are not taken into consideration from the general WiFi technology. As the number of these devices increases, the 2.4 GHz and 5 GHz channels become overcrowded while the interference increases with a drop in the network capacity. Unfortunately, traditional Wi-Fi was not originally designed to tackle these interference issues and the increasing capacity in dense environments. To fill this gap, WiFi Alliance announced the WiFi HaLow (IEEE 802.11 ah).

\item \textbf{IEEE 802.11ah - WiFi HaLow.} WiFi HaLow \cite{halow} was designed to enable connectivity to  a variety of new, power-efficient use cases in smart homes and smart cities, connected vehicles and supporting the concept of the IoT in general. It extends the WiFi into the 900 MHz band to enable low power connectivity which is necessary for the IoT devices. The transmission range is twice the range of WiFi while it increases the signal robustness  in challenging environments, such as complex indoor environments with lots of furniture  and walls. It can operate in multiple transmission modes from low-rate starting from 150 Kbp/s and up to 347 Kbp/s.
  
The ability to operate in low power, the high transmission range and the low propagation loss makes WiFi HaLow a good candidate for micro-location with IoT devices. On the other hand, it is relatively new in comparison with other technologies, published in 2017, hence it is not widely available and it will be a while before we see HaLow clients and infrastructure device. This delays the experimentation that is necessary before deciding if it is suitable for micro-location.

\item \textbf{Zigbee.} Zigbee is a high-level communication protocol known for its simplicity, low-power usage, and secure networking \cite{zigbee}. It is based on the IEEE 802.15.4 standard, which defines the operating point of Wireless Personal Area Networks (WPANs) with low data rate antennas. They are able to control the flow of information and prevent any loss of data by using Carrier-Sense Multiple Access with Collision Avoidance (CSMA/CA). Devices using Zigbee are designed with additional features such as link quality and energy detection, which allow for measurements such as the RSSI to be easily determined. 
 
Zigbee is commonly used for localization in WSN due to its low power requirements. Among the IoT devices through it is not popular due to the extra hardware that is needed.

\item \textbf{Bluetooth. } Bluetooth is  another wireless technology for  exchanging data over short distances \cite{blue5}. The IEEE standardized Bluetooth as IEEE 802.15.1, but no longer maintains the standard, which is managed by the Bluetooth Special Interest Group (SIG). According to the Group, Bluetooth is all about proximity, not about exact location. Bluetooth was not intended to offer a pinned location like GPS. However is known as a geo-fence or micro-fence solution which makes it an indoor proximity solution, not an indoor positioning solution. 

Introduced by Bluetooth Special Interest Group in 2010,  Bluetooth Low Energy (BLE) was designed for applications that do not require large amounts of data transfer, while reducing the power consumption and cost of devices. Micro-location and indoor mapping have been linked to Bluetooth and to the BLE-based iBeacon promoted by Apple Inc. \cite{ibeaconpacket}. Large-scale indoor positioning systems based on iBeacons have been implemented and applied in practice.

Similar to Zigbee, BLE is a technology used in WPANs. The low power consumption of BLE has led to a number of new devices in the IoT.  BLE 4.0 can reach 25 Mbit/s at a distance of 60 m. Applications utilizing BLE have greatly increased over the past couple of years. A number of new devices have been developed, ranging in fields such as healthcare~\cite{kennedy}, sports, fitness, security, and home entertainment. One device that has been created is known as a beacon. Beacons are small, inexpensive devices, which contain only a CPU, a radio, and batteries. 
 
Bluetooth 5.0 \cite{blue5} is the competitor of WiFi HaLow in the IoT domain. It is claimed to have twice the speed of the previous  version, four times longer transmission range, and exchange data eight times faster. 

The simplicity and the popularity among IoT devices are advantages of Bluetooth for micro-location. The small size of beacons and their low cost with the energy efficiency of the BLE and the extended lifespan that it can provide, can be used to enhance micro-location in a complex environment without interfering with other wireless infrastructures. On the disadvantages, even though the security of BLE is good, it is even better on Wi-Fi.

\item \textbf{RFID.} Radio Frequency Identification Device (RFID) was primarily designed for data transferring and storing \cite{rfid}. There is a need for an RFID reader that can communicate with the RFID tags. There are two types of RFIDS. The Active RFIDs operate in the Ultra High Frequency (UHF) and microwave frequency range. They need to be connected to a local power source while the transmit their ID periodically up 100 m. On the other hand, Passive RFIDs, they operate without battery but within 1-2 m transmission range.

In the IoT era, RFID is not a promising solution for micro-location. Their accuracy is not high enough while they are not available on many portable devices.

\item \textbf{LoRaWAN.} LoRaWAN is a long-range, low power consumption technology used in the development of Personal Wide Area Networks (PWAN) \cite{lorawan}. Originally developed by the LoRa Alliance, the LoRaWAN protocol transmits at a lower frequency of 915MHz. The benefit of using a lower frequency is that the smaller wavelength allows for a greater distance that the signal can reach. Due to that, it can pass through walls and obstacles without issue. It is also no longer as easily susceptible to noise since it does not interfere with any devices transmitting on the 2.4GHz band. 

The disadvantage of using such a low frequency is a reduction in the data rate that can be sent between transmitting devices. For micro-location, this is not an issue as the nodes are not transmitting large amounts of information. Due to the 915MHz band being unlicensed it is free for anyone to use for their personal networking needs. 
 
For devices that are moving at high speed in a large area, LoRa might be a candidate for localization with IoT. Unfortunately, in short range, LoRa performance does not overcome the high cost and the extra equipment that is needed to set up a LoRa node.

\item \textbf{LiFi:} LiFi is a VLC technology \cite{Lifi}.  VLC is a subset of Optical Wireless Communication (OWC), which utilizes the light emitting diodes as a medium to enable high speed communication. Data is transmitted by modulating the intensity of LED light at nanosecond intervals, too quick to be detected by the human eye. 
    
 \end{itemize}
   
Table \ref{wirelesstech} summarizes the specifications of each wireless technology along with the advantages and disadvantages of usage for micro-location.
    
\begin{table}[t!] 
\begin{center}
\begin{tabular}{ |l| c|c|c|p{2.8cm}|p{2.8cm}| }
\hline
 \textbf{Technology} & \textbf{Throughput} & \textbf{Transmission Range}  & \textbf{Power Consumption} & \textbf{Advantages} & \textbf{Disadvantages}\\ 
 \hline \hline
 \textbf{IEEE 802.11ac}  & 3.5 Gbit/s &  35 m  &  \multirow{2}{*}{Moderate} & \multirow{2}{2.8cm}{Availability in many environments} &  \multirow{2}{2.8cm}{Prone to noise and interference}  \\   
 \textbf{IEEE 802.11ad}  & 6.7 Gbit/s &  3.3 m  &   & &\\   \hline

 \textbf{IEEE 802.11ah}  & 347 Mbit/s&  1 km  & Low & Wide reception range & Not widely available  \\   \hline
  \textbf{ZigBee} & 250 kbit/s & 75 m & Low  & Easy to setup & Extra hardware\\  
 \hline
  \textbf{BLE v4.0} &  25 Mbit/s & 60 m &   \multirow{2}{*}{Low} & \multirow{2}{*}{High throughput} & \multirow{2}{*}{Prone to interference}  \\    
  \textbf{BLE v5.0} &  50 Mbit/s & 240 m &   &  &   \\    \hline

  \textbf{RFID Active} & 1067 & 100 m &  \multirow{2}{*}{Low} & \multirow{2}{*}{Low power}&\multirow{2}{*}{Low accuracy}    \\  
  \textbf{RFID Passive} &1067 & 2 m &  & &     \\  \hline

 \textbf{LoRaWAN} & 50kbps & 15 km & Extremely Low & Wide  range& Extra hardware  \\  \hline
   \textbf{LiFi} & 1Gbps & 10m  & Low & Dense Environments & Low Range     \\  \hline

\end{tabular}
\end{center}
\caption{Wireless Technologies for micro-location.}
\label{wirelesstech}
 
\end{table}

\subsection{Radio Signal Features for Micro-location.}
As the wireless signal propagates from the sender to the receiver, there are signal characteristics that can be used for the localization  of one of the communicating devices. There are four main signal features that can be used for localization:

\begin{itemize}

\item \textbf{Received Signal Strength Indication (RSSI).}
Received Signal Strength Indication (RSSI)  is one of the most commonly used characteristics for indoor localization \cite{sadowski}. It is based on measuring the power present in a received signal from a client device to an access point. As radio waves propagate according to the inverse-square law, the distance can be approximated based on the relationship between transmitted and received signal strength, as long as no other errors contribute to faulty results.  The combination of this information with a propagation model can help to determine the distance between the client device and the access points. Lateration-based methods are commonly used along with RSSI to estimate the location of the client.

It can be assumed that the more access points, the more information can be collected hence, the accuracy can be increased. This, however, works also as a trade-off. An increase of the access points will also increase the interference between different signal. A key challenge in wireless localization systems is that the range measurements are often associated with errors. Although RSSI techniques are among the cheapest and easiest methods to implement, its disadvantage is that it does not provide very good accuracy, with a median of 2 to 4 m. This is mainly because the RSSI measurements tend to fluctuate according to environmental changes or multipath fading, events that are common in indoor environments.

\item \textbf{Angle of Arrival (AoA).} Angle of arrival (AoA) is another characteristic that can be used for localization. It tries to estimate  the direction of the signal propagation, i.e. the  angle from which the signal arrives at a receiver.  AoA is typically achieved by using an array of antennas.  The line connecting two reference points may be used as an internal reference. The spatial separation of antennas leads to differences in arrival times, amplitudes, and phases.

\item \textbf{Time of Arrival (ToA).} In Time of Arrival (also know as time of flight), the distance between sender and receiver of a signal can be determined using the measured signal propagation time and the known signal velocity. ToA is the amount of time a signal takes to propagate from transmitter to receiver. The signal propagation rate is constant and known hence,  the travel time of a signal can be used to directly calculate distance. This is the technique used by GPS.

The accuracy of the TOA-based methods often suffers from massive multipath conditions in indoor localization, which is caused by the reflection and diffraction of the RF signal from objects (e.g., interior wall, doors or furniture) in the environment. However, it is possible to reduce the effect of multipath by applying temporal or spatial sparsity-based techniques

\item \textbf{Time Difference of Arrival (TDoA).}  The Time Difference of Arrival (TDoA) is  the time of arrival of a specific signal, at physically separate receiving stations with precisely synchronized time references.  TDoA measures the difference in ToA at two different receivers. Three or more TDoA measurements can be used to locate a device using hyperbolic
lateration. 

Although TDoA sounds similar with ToA, there is a difference. In ToA the absolute time at a base station is used. In TDoA the measured time difference between departing from one and arriving at the other station is used.

\end{itemize}

\subsection{Indoor Positioning Techniques.}
In this section some commonly used techniques for localization are discussed.
\begin{itemize}
\item \textbf{Proximity Detection.} Proximity detection techniques, shown in Fig. \ref{proximity}, are based on the proximity of the mobile device to previously known locations. These techniques determine the position of an object based on closeness to a reference in the physical space. When the mobile device receives the signal from a reference point, then the device should be in within the coverage range of the reference point, i.e. in close proximity to the reference point. Proximity detection does not provide the location in form of coordinates but rather in form of sets of possible locations. 
 
 This method also based on the premise that the reference point has a limited range. For simplicity, it is common to assume that the range of a wireless infrastructure would be well represented by a circle of given radius $r$. Then, the result of the proximity detection  in being located inside this circle. For several circles, one can limit the possible location to the intersection of the different circles.

\item \textbf{Lateration.} Lateration is the process of estimating the location of a mobile device given distance measurements to a set of points with a known location, shown in Fig. \ref{lateration}. Lateration-based methods use the distance measurements from multiple reference points to compute the position of a receiver.  Trilateration  is a commonly used technique to calculate the estimated client device position relative to the known position of  three access points. It uses the distance from the three reference points to estimate the location and  track the position of the receiver when the receiver is moving within the three points.  Given the distance to an anchor, it is known that the node must be along the circumference of a circle centred at anchor and a radius equal to the node-anchor distance. In two-dimensional space, at least three non-collinear anchors are needed and in three-dimensional space, at least four non-coplanar anchors are needed. 
 
\item \textbf{Angulation.} Angulation-based positioning techniques can be used to employ the angle of arrival of a wireless signal and determine the position of a receiver, shown in Fig. \ref{angulation}. A commonly used approach is triangulation were  the location of a point is determined by forming triangles to it from known points. In triangulation, a known baseline can  be used  to find the location relative to two anchor transmitters. It uses the geometric properties of triangles to estimate the location and relies on angle (bearing) measurements. It requires a minimum of two bearing lines and the locations of anchor nodes or the distance between them, for two-dimensional space.

\item \textbf{Fingerprinting.}  Fingerprinting techniques are based on the reproducibility of patterns of measurable variables, shown in Fig. \ref{finger}. Traditional fingerprinting
 records the signal strength from several access points  and store them in a database, along with the known coordinates of the client device in an offline phase. Then, during the localization phase, the current vectors at an unknown location are compared to those in the database and the closest match is returned as the estimated user location.
 
Fingerprinting has the advantage that it does not require any assumption regarding the nature of the propagation environment. It just creates a modelled environment based on the training data. At the same time, this can be a disadvantage. Any change of the environment, such as adding or removing furniture or access points, require an update to the model.

\end{itemize}

\newcommand{\gscale}{0.2}
\begin{figure*}[t!]
\centering
\captionsetup[subfloat]{farskip=0pt}%
\subfloat[Proximity.]{\includegraphics[scale=0.2]{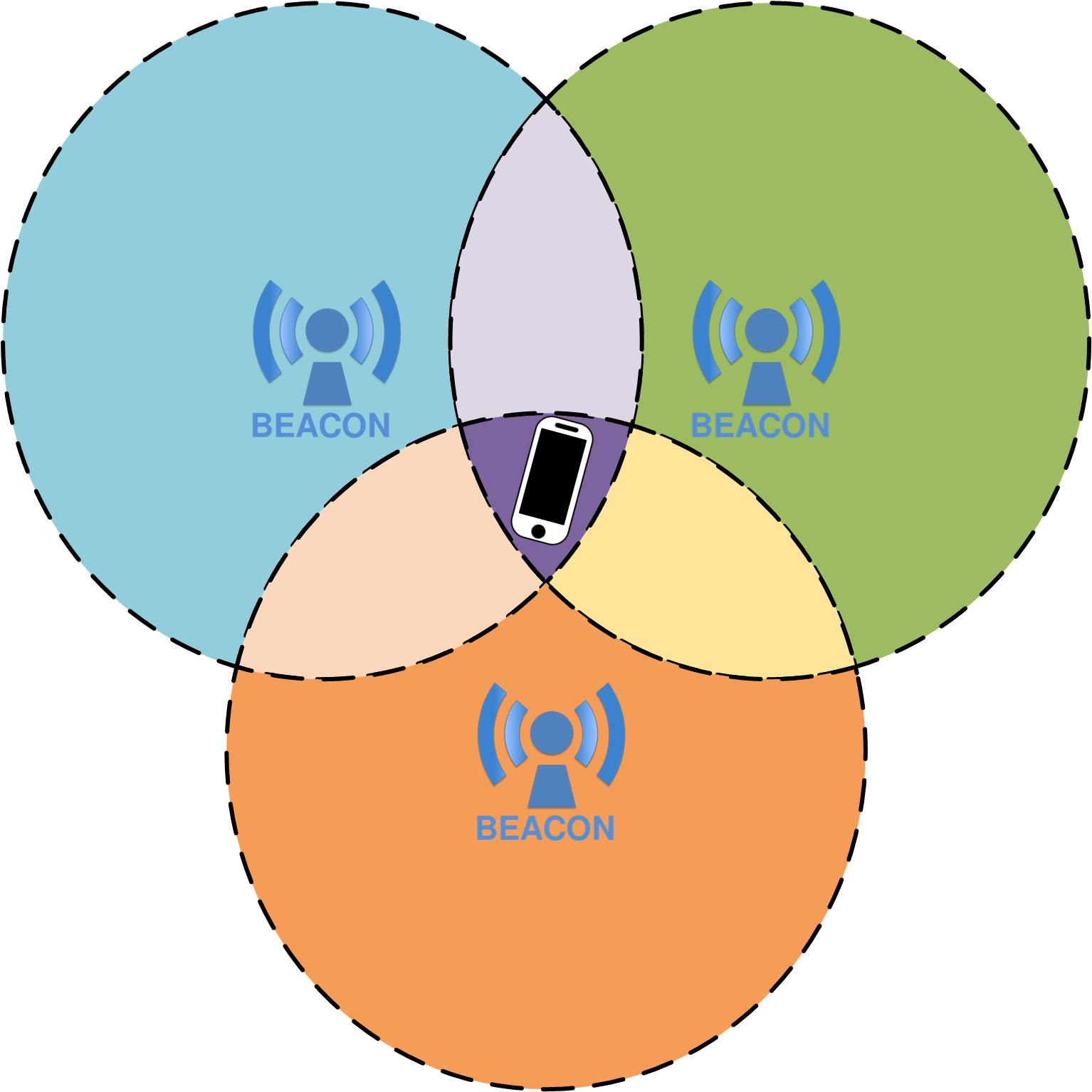}
\label{proximity}}\hfill
\subfloat[Lateration.]{\includegraphics[scale=\gscale]{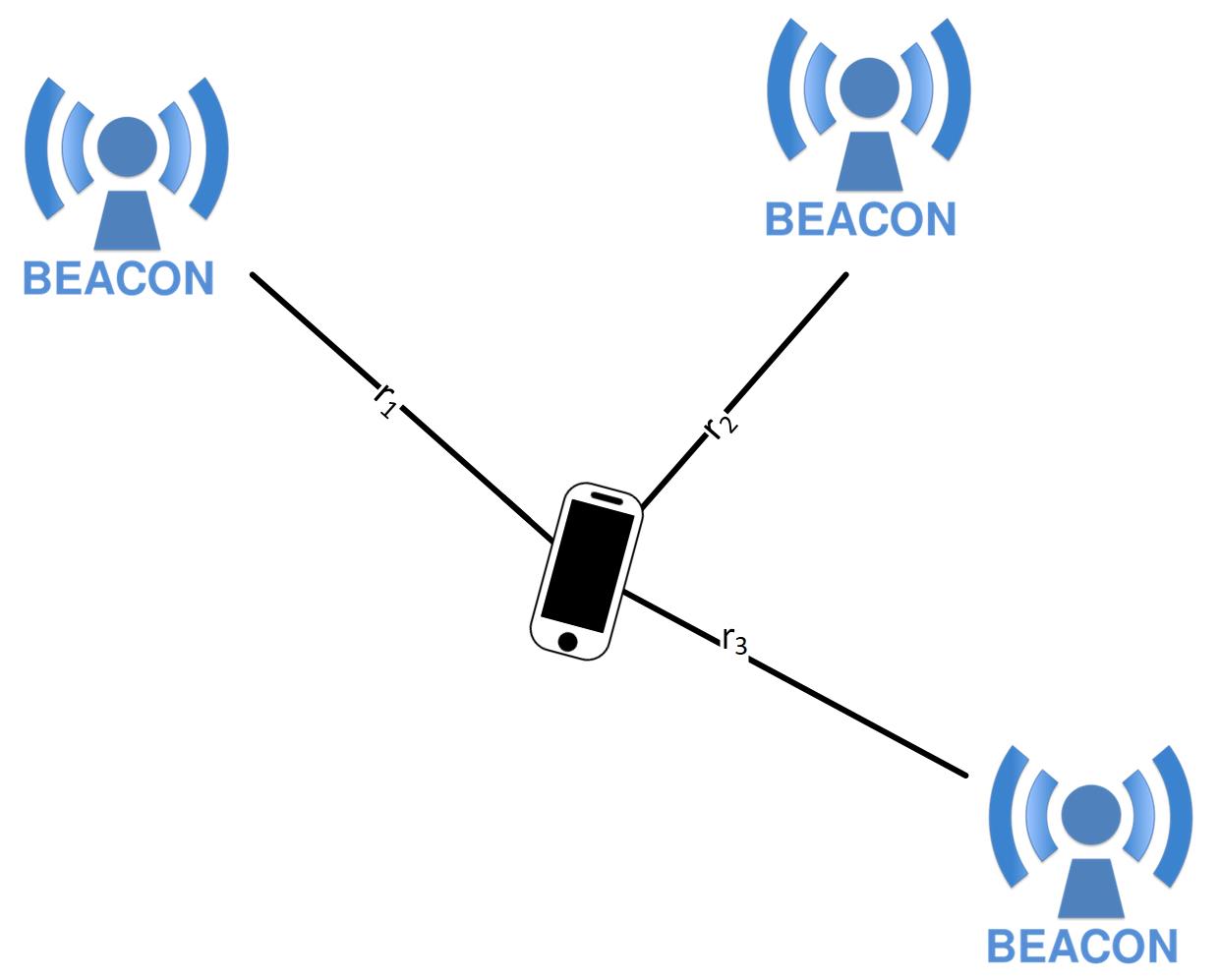}
\label{lateration}}\hfill
\subfloat[Angulation.]{\includegraphics[scale=\gscale]{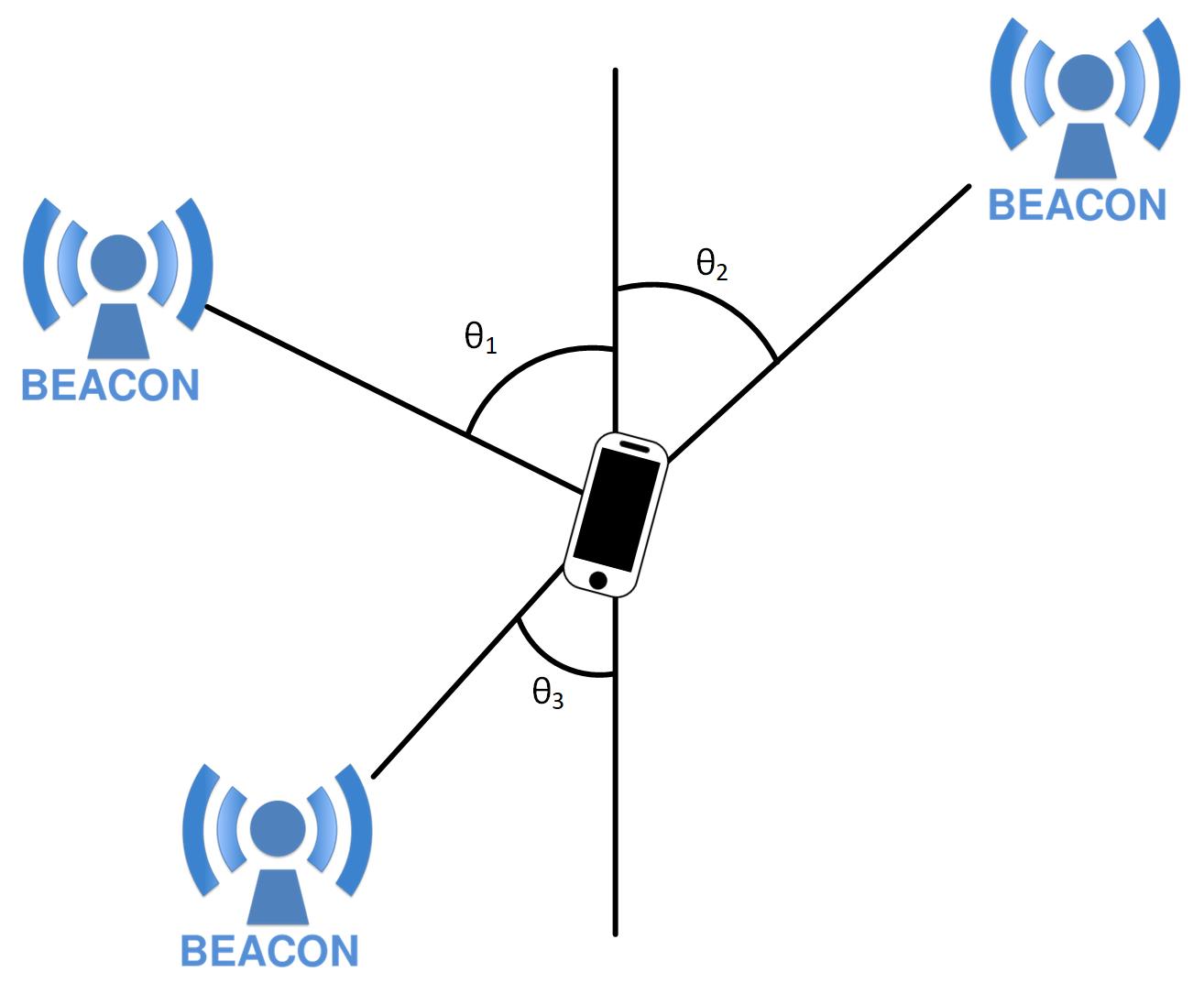}%
\label{angulation}}\hfill
\subfloat[Fingerprinting.]{\includegraphics[scale=\gscale]{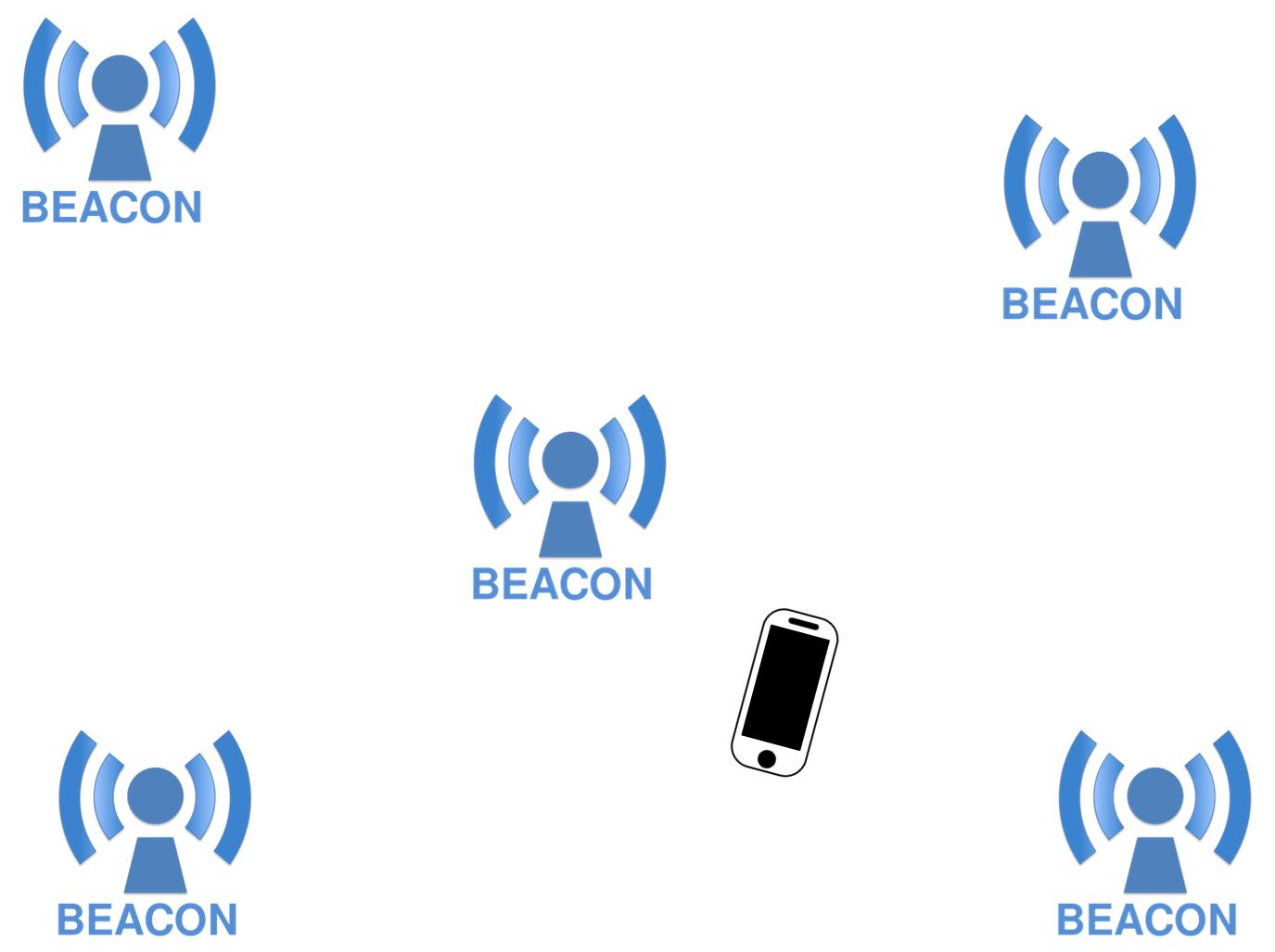}%
\label{finger}}

\caption{Localization Techniques.}
\label{techniques}
\end{figure*}

\subsection{Localization Metrics.}
To evaluate the performance of a localization system, accuracy and precision are used. Accuracy measures the deviation of the estimated location from the truth, while precision measures the deviation of location estimates from each other for the same location. A system with high accuracy can be used for an application that focuses on the long-term localization determination and the errors cancel out over time. On the other hand, a system with high precision can be used to find the proximity between devices but it is hard to be used for localization.

\section{Improve Accuracy through Signal Processing Filtering Techniques.}
There are a number of signal processing filtering techniques that are used for indoor localization. In the following, we summarize two: Kalman Filtering (KF-ST) and Dynamic Kalman Filtering (KF-DN).
 
\subsection{Indoor Localization Model.}
We model the indoor localization problem as posed by Arulampalam et al. \cite{arulampalam2002tutorial}. Extended versions as applied in BLE can also be found in \cite{zafari2017ibeacon}. Since we seek to estimate the user position/state under a set of measurements obtained in a typical noisy indoor environment, Bayesian filtering is an attractive approach for such problems. However,  Bayesian filtering requires the following two models.

\begin{enumerate}
	\item \textbf{System Model}: A system model describes the variation of the state (user position in our case) with time.  The system model relates the position vector $y_i$ with the process noise $m_i$ and previous state.
	\item \textbf{Measurement Model}: A measurement model relates the noisy measurements (RSSI for PF and the user position for EKF) with the state/position.
\end{enumerate}
We construct the Posterior Probability Density Function (PDF) describing the state from all available information, including the measurements from the reference nodes (beacons in our case). The PDF is considered as the complete solution to the state estimation problem, since it contains all the required information. The problem involves recursively estimating the user state/position as we receive measurements from the beacon. Therefore, we require a recursive filter. Recursive filters consist of the prediction and update stage in which the state is predicted and then updated once the measurements are available. The presence of noise in  indoor settings affects the position calculation so the pdf is usually distorted. The obtained measurements in the update state are used to modify the prediction pdf using Bayes theorem. 
\par Mathematically,   state $y_i$ at time $i$ is a function of the state at time step ($i-1$) as well as the process noise $m_{i-1}$ \cite{djuric2003particle} as described in Equation \eqref{eq:sysmodel}:
\begin{equation}
\protect\label{eq:sysmodel}
y_i=f_i(y_{i-1}, m_{i-1})
\end{equation}

$f_i: \Re^{n_y}$ x $\Re^{n_m}  \rightarrow \Re^{n_y}$ is the non-linear function (as indoor localization is a non-linear problem) that relates the previous state $y_{i-1}$ and process noise $m_{i-1}$ with the current state $y_i$ as described by Arulampalam \cite{arulampalam2002tutorial}.  The sequence $\{m_{i}, i\in\aleph\}$ represents an independent, identically distributed (i.i.d) process noise sequence.  The integer ${n_y}$ represents the state noise vector, and ${n_m}$ represents the process noise vector. $\aleph$ represents the set of Natural numbers. The measurement model relates the obtained measurement $x_i$ to the state $y$ and measurement noise $n$ at time $i$ \cite{djuric2003particle} as given in Equation \eqref{eq:measmodel}:
\begin{equation}
\protect\label{eq:measmodel}
x_i=h_i(y_{i}, n_{i})
\end{equation}
The mapping function $h_i: \Re^{n_y}$ x $\Re^{n_n}  \rightarrow \Re^{n_x}$  can be either linear or non-linear.  Functions $f_i$ and $h_i$ rely on the laws of motion/physics. 
 The sequence $\{n_{i}, i\in\aleph\}$ is a measurement noise sequence that is independent and identically distributed. The integers ${n_x}$ and ${n_n}$ represent the  measurement and measurement noise vectors dimension respectively.
\par Recursively calculating the pdf $p(y_i|x_{1:i})$ allows us to continuously calculate the belief in the state $y_i$ at any particular time instance $i$ in the presence of noisy measurements. The initial pdf  $p(y_o|x_0)$ is assumed to be equivalent to state vector's prior $p(y_0)$ \cite{arulampalam2002tutorial}. We assume that the prior is available. The available information is enough to calculate the pdf $p(y_i|x_{1:i})$ recursively in the prediction and update stages. In the prediction stage if the pdf $p(y_{i-1}|x_{1:i-1})$ is available,  we  can use  Chapman-Kolmogorov equation given in  Equation \eqref{eq:chapman} to obtain the  prior pdf of the state at any time instance $i$.
\begin{equation}
\protect\label{eq:chapman}
p(y_i|x_{1:i-1}) = \int p(y_i|y_{i-1}) p(y_{k-1}|x_{1:i-1})dy_{i-1}
\end{equation}
At any time instance  $i$,  we collect the observations $x_i$ from the sensors to update the prior using Bayes rule given in Equation \eqref{eq:meas} \cite{arulampalam2002tutorial}. The denominator in Equation \eqref{eq:meas} is explained in Equation  \eqref{eq:normal}.
\begin{equation}
\protect\label{eq:meas}
p(y_i|x_{1:i}) =  \frac{p(x_i|y_i)p(y_i|x_{1:i-1})}{p(x_i|x_{i-1})}
\end{equation}

\begin{equation}
\protect\label{eq:normal}
p(x_i|x_{i-1})= \int {p(x_i|y_i)p(y_i|x_{i-1})dy_i}
\end{equation}
The collected measurements $x_i$ in the update stage are then used  to update the prior density, resulting in the required current state's posterior density.  Recursively updating the system using Equations \eqref{eq:chapman}$ $ and \eqref{eq:meas} result in an optimal Bayesian solution. However analytically, it is not possible to obtain the recursive propagation of posterior probability density as done in Equations \eqref{eq:chapman}$ $ and \eqref{eq:meas}. Therefore, a number of different algorithms including PF, KF and EKF are used to obtain a solution.
  
\subsection{Kalman filter.}
The Kalman filter based RSSI smoother is based on the work of Guvenc \cite{guveenhancements}. 
The state $x_i$ that in our case consists of RSSI and rate of change of RSSI, at time $i$ is a  function of the state at time $i-1$ and the process noise $w_{i-1}$ which is given mathematically by Equation \ref{eq:1}. The obtained RSSI measurements $z_i$ at instant $i$ from the iBeacons is a  function of the state at $i-1$ and the measurement noise $v_i$ as given by Equation \ref{eq:2} as described in Arulampalam \cite{arulampalam2002tutorial}.
\begin{equation}
\label{eq:1}
x_i=f(x_{i-1},w_{i-1})
\end{equation}
\begin{equation}
\label{eq:2}
z_i=h(x_{i-1},v_i)
\end{equation}
The traditional Bayesian based approach consists of the prediction and update stage as described by Guvenc \cite{guveenhancements} and is given below:
\begin{enumerate}
	\item Prediction Stage
	\begin{equation}
	\label{eq:3}
	p(x_i|z_{1:i-1})=\int p(x_i|x_{i-1})p(x_{i-1}|z_{1:i-1})dx_{i-1}
	\end{equation}
	\item Update Stage:
	\begin{eqnarray}
	\label{eq:4}
	p(x_i|z_{1:i})= & \frac{p(z_i|x_i)p(x_i|z_{1:i-1})}{p(z_i|z_{1:i-1})}
	\end{eqnarray}
	where
	\begin{eqnarray}
	\label{eq:5}
	p(z_i|z_{1:i-1}) = & \int p(z_i|x_i)p(x_i|z_{1:i-1})dx_i
	\end{eqnarray}
\end{enumerate}
We assume that both the process noise and measurement noise are Gaussian and the functions $f$ and $h$ in Equations \ref{eq:1} and \ref{eq:2} are linear. Because of the linearity assumption, we can apply a Kalman filter since it is the optimal linear filter. 

Due to the aforementioned assumptions, Equations \ref{eq:1} and \ref{eq:2} can be rewritten as described by Guvenc \cite{guveenhancements}
\begin{equation}
\label{eq:6}
x_i=Fx_{i-1}+w_i
\end{equation} 
\begin{equation}
\label{eq:7}
z_i=Hx_i+v_i
\end{equation}
where $w_i \sim N(0,Q)$ and $v_i \sim N(0,R)$. Table  \ref{tab:1} lists the parameters of a Kalman filter.
\begin{table}
	\centering
	\caption{Kalman filter parameter notation}
	\label{tab:1}
	\begin{tabular}{|l| p{6cm}|}
		\hline
		\textbf{Symbol} & \textbf{Meaning}                 \\ \hline
		x      & State vector                     \\ \hline
		z      & Measurement/observation vector   \\ \hline
		F      & State transition matrix          \\ \hline
		P      & State vector estimate covariance or Error covariance \\ \hline
		Q      & Process noise covariance         \\ \hline
		R      & Measurement noise covariance     \\ \hline
		H     & Observation matrix               \\ \hline
		K & Kalman Gain \\\hline
		w & Process noise  \\ \hline
		v & Measurement noise \\ \hline
	\end{tabular}
		\vspace{-12pt}
\end{table}
The prediction and update stages for the Kalman filter as described by Guvenc \cite{guveenhancements} are
\begin{enumerate}
	\item Prediction Stage: 
	\begin{equation}
	\label{eq:8}
	\hat{x}_{\bar{i}}= F\hat{x}_i
	\end{equation}
	\begin{equation}
	\label{eq:9}
	P_{\bar{i}}= FP_{i-1}F^T+Q
	\end{equation}
	\item Update State:
		\begin{eqnarray}
	K_i=& P_{\bar{i}}H^T(HP_{\bar{i}}H^T+R)^{-1}\\
	\hat{x}_i=&\hat{x}_{\bar{i}}+K_i(z_i-H\hat{x}_{\bar{i}})\\
	P_i= &(I-K_iH)P_{\bar{i}}
	\end{eqnarray}
\end{enumerate}
The higher the Kalman gain, the higher will be the influence of the measurements on the state. The prediction and update steps are recursive in nature. 
For the purpose of filtering the RSSI values, we utilize a state vector $x_i$ that consists of the RSSI value $y_i$ and the rate of change of RSSI $\Delta y_{i-1}$ as given below. 
\begin{equation*}
x_i=\begin{bmatrix}
y_i\\ 
\Delta y_i
\end{bmatrix}
\end{equation*}
$\Delta y_i$ is dependent on the environment and signifies how drastically RSSI value fluctuates. The higher the noise in the environment, the higher will be the fluctuation. The current value of RSSI $y_i$  is assumed to be the previous RSSI $y_{i-1}$  plus the change $\Delta y_i$ and process noise $w_i^y$. Hence Equation \ref{eq:6} can be written as 
\begin{equation}
\label{eq:13}
\begin{bmatrix}
y_i\\ 
\Delta y_i
\end{bmatrix}
= \begin{bmatrix}
1 & \delta t \\ 
0 & 1 
\end{bmatrix}
\begin{bmatrix}
y_{i-1}\\ 
\Delta y_{i-1}
\end{bmatrix}
+
\begin{bmatrix}
w_i^{y}\\ 
w_i^{\Delta y}
\end{bmatrix} 
\end{equation}
which means that the state transition matrix F is given by 
\begin{equation*}
F=\begin{bmatrix}
1 &\delta t \\ 
0 & 1 
\end{bmatrix}
\end{equation*}
The parameter $\delta t$ is to be adjusted as per the variation in RSSI which depends on the environment. For our set of experiments,  $\delta t$ was taken as 0.2 (using trial and error). Similarly, Equation \ref{eq:7} can be rewritten as 
\begin{equation}
\label{eq:14}
\begin{bmatrix}
z_i
\end{bmatrix}
= \begin{bmatrix}
1 & 0  

\end{bmatrix}
\begin{bmatrix}
y_{i}\\ 
\Delta y_{i}
\end{bmatrix}
+
\begin{bmatrix}
v_i^{y}
\end{bmatrix} 
\end{equation}
The observation matrix H is given by 
\begin{equation*}
H=\begin{bmatrix}
1 & 0  
\end{bmatrix}
\end{equation*}
Parameters P, Q and R used in the experiments were obtained using trial and error, and are given below.
\begin{equation*}
\label{eq:15}
P = 100\textbf{I}_{22}, \;\; 
Q = 0.001\textbf{I}_{22}, \;\; 
R =
\begin{bmatrix}
0.10 
\end{bmatrix} 
\end{equation*}
The Kalman filter, once calibrated, effectively smooths the RSSI values. The smoothed RSSI values were then input into the path-loss model to obtain distances between the iBeacons and the user, and the user's proximity to the beacon was classified in any of the aforementioned zones.

\subsection{Dynamic Kalman.}
A dynamic variation of the  Kalman filter computes $Q$ as the variance of a set number of previously collected RSSI values to make up for real-world process noise changes. It is continuously recalculated at each iteration of reading in the next RSSI value.

Different set sizes of recorded RSSI values can be used to find the ideal number of values to utilize in this calculation.  It can be inferred that as the array size increases the accuracy increases as well, up to an array size of $n$.  After a size of $n$, any increase of the size leads to the decrease of the accuracy. The optimal $n$ can be found through experimentation while the increase of the size can lead to waste of resources without any increase in the accuracy. At the same time, a decrease of the size below $n$ does not give sufficient information to the system to increase its prediction accuracy.

The set of RSSI values is stored in an array list of data type double. Algorithm ~\ref{maint} illustrates the procedure. It adds entries to each index in increasing order starting from index 0. When an entry is deleted, all entries ahead get pushed down one index value. At the start of each iteration the algorithm checks the size of the array, and once it reaches the desired size $n$, it removes the oldest entry (index 0) and adds in the newest measurement.\\

\begin{algorithm}
\caption{Maintain RSSI Set.}
\begin{algorithmic}[1]

\If {$RSSI Array.size() == n$}
\State $remove RSSI Array[0]$.
\EndIf
\State $lastIndex \gets RSSI Array.size()$ 
\State $RSSI Array[lastIndex] \gets \textit{newRSSI}$

\end{algorithmic}
\label{maint}
\end{algorithm}

When developing the dynamic noise component of the Kalman filter, it is essential to find the ideal number of previously obtained RSSI values to maintain, for calculation purposes. This is because the size of this set will have a direct impact on the performance of the filter.

\section{BLE Beacon Technology.}

In this section, an introduction to BLE beacons is presented along with their characteristics and their most commonly used wireless protocols.

\subsection{BLE beacons.}

\begin{figure}[t]
\centering%
\includegraphics[width=0.4\columnwidth]{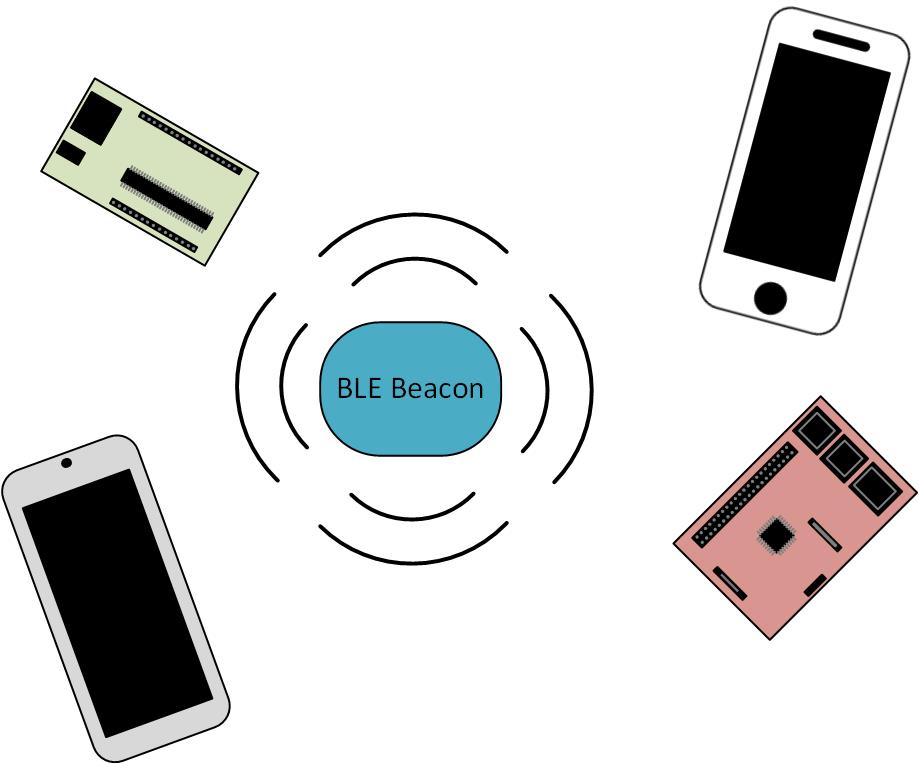}%
\caption{A BLE beacon broadcasting a signal to the nearby devices. Each device can receive the signal and take an action in response.}%
\label{beacontrans}%
\end{figure}

BLE beacons are small wireless transmitters that broadcast their identifier to nearby electronic devices, such as smartphones, wearables, and other IoT devices. An analogy of the way beacons work is with the operation of a lighthouse. The lighthouse represents a known location which can be uniquely identified by its light. All the ships that can see the light, they know about the existence of the lighthouse. On the other hand, the lighthouse neither communicates with the ships nor does it knows how many ships see its light or how many other lighthouses are in the area. Similarly, every beacon is sending out a radio signal to inform all the radio enable devices in its range that the beacon is there. It does not know how many beacons or receiving devices are in the area and  it does not connect with them. An example of beacon operation is shown in Fig. \ref{beacontrans}.

Beacons  broadcast signals at a certain interval and within a certain transmission range.  A beacon broadcasts a signal to all nearby devices that can receive the Bluetooth signal, i.e. the devices that have a Bluetooth receiver and the receiver is on.  In order to collect the signal from the beacon, it is necessary to have a device with a BLE receiver. This can be a smartphone or a single-board computer like Raspberry Pi. Applications or functions can be implemented  based on the  signal from the beacons. However, these applications are running on the hosting device, i.e. a smartphone or a Raspberry Pi, and not on the beacon.

Beacons are using BLE.  The way how the peripheral device announces its existence to the other devices is the opposite of how it is in the original Bluetooth Classic.  BLE enables a peripheral device to transmit an advertisement packet without being paged by the master/ central device \cite{kriz}. Due to this communication model, it is possible to construct energy-efficient transmitters. Moreover, when two BLE 4.0 devices are paired, they waste less battery power because the connection is dormant unless critical data is being shared. With the previous generation of Bluetooth, it was best to shut down your hardware when it was not in use.  the Bluetooth Special Interest Group estimates between 1 and 2 years of battery power in some devices with Bluetooth 4.0.

\subsection{Configuration parameters.}

BLE beacons have  configuration parameters and a set of values that can determine their performance and their utility for different applications. Some of these parameters are important when beacons are used in micro-location applications.

\begin{itemize}

\item \textbf{Transmission power.} Transmission power is the required power to broadcast the beacon signal. As in every wireless device, transmission power directly affects the transmission range. The higher the transmission power, the longer the signal range of the beacon.  This is an important trade-off for most of the beacon applications. Technically, a beacon's range can reach up to 70 m however, the battery might only last for six months. If the transmission range is constrained to 2 m, then the beacon might go up to two years without the need of battery replacement.  A small transmission power  can also increase the required number of beacons to cover an area while a large transmission power can increase the collisions and the interference. As it can be inferred, an optimal transmission range can help to extend the lifetime of the beacons and minimize the battery replacement cost. At the same time, it can minimize unnecessary collisions with other beacons in the area. 

\item \textbf{Advertising interval.} Advertising interval is another characteristic that affects the overall performance of beacons. It describes the time between consecutive transmissions.  Applications that need to notify or detect the users that are moving in the area, they need a short advertising interval while  applications where the users are moving less frequently, they  might improve their performance with a longer advertising interval. Similar to the transmission power, the advertising interval affects beacons performance. The shorter the interval the more stable the signal from the beacon. At the same time, the shorter the interval the higher the power consumption. Once again, there is a trade-off between beacon performance and power consumption.

\item \textbf{BLE beacon protocols.}
Beacon protocols are standards of BLE communication. Each protocol describes the structure of the advertisement packet beacons broadcast. It is necessary for the advertisement packet to have the MAC address of the beacon. There are different protocols, the most popular of which are the following:

\begin{itemize}

\item \textbf{iBeacon}. Apple's iBeacon was the first BLE Beacon technology to come out \cite{ibeaconpacket}. iBeacon is a proprietary, closed standard. It broadcasts four pieces of information:

\begin{enumerate}

\item A Universally Unique IDentifier (UUID) that identifies the beacon.

\item A Major number identifying a subset of beacons within a large group.

\item A Minor number identifying a specific beacon within the subset.

\item A transmission power level in 2's complement, indicating the signal strength one meter from the device. This number must be calibrated for each device by the user or manufacturer.
\end{enumerate}

iBeacon has a simple implementation and large documentation but it has fewer features in comparison with the following protocols. iBeacon works with iOS and Android, but are native to iOS.

\item \textbf{Eddystone}. Eddystone was announced from Google and it is another protocol that defines a BLE message format for proximity beacon messages \cite{eddystonepacket}. Eddystone protocol is able to transmit four different frame-types:
\begin{enumerate}
\item UID which is used to identify the individual beacon.
\item URL which can be  a website link that redirects to a website that is secured using SSL, eliminating the need for a mobile app.
\item TLM which includes sensor and administrative data from the beacon through telemetry. Examples include the beacon's battery level and its temperature.
\item EID which is an encrypted ephemeral identifier that changes periodically at a rate determined during the initial registration with a web service. This frame type is intended for use in security and privacy-enhanced devices.
\end{enumerate}
Eddystone also works with both iOS and Android.

\item \textbf{AltBeacon}.  It is an open source beacon protocol \cite{altbeacons} that was designed by Radius Networks. It has the same functionality as an iBeacon but it is not company-specific. This makes AltBeacon compatible with any mobile operating platform and more flexible since they have customizable source code.

\item \textbf{GeoBeacon}. It is another open source beacon protocol, designed for usage in GeoCaching applications \cite{geobeacon}. It has a very compact type of data storage. GeoBeacon can provide high-resolution coordinates and it is also compatible with different mobile operating platforms.
\end{itemize}

\end{itemize}

\subsection{Hardware Solutions}
There is a great variety of BLE beacon devices on the market, most of them operate on batteries such as Estimote, Kontakt, Gimbal, Glimworm, BlueCats and more \cite{aislelabs}, while there are also solar power beacons like the CYALKIT- E02. Each has its own unique features, such as additional sensors, battery life, reconfigurability, and dimensions, though all fundamentally work the same. 

At the physical layer, BLE transmits in the 2.4 GHz Industrial,
Scientific, and Medical (ISM) band with 40 channels, each 2.0 MHz wide. From those channels, 37 are used to exchange the data among paired devices and 3 channels are designated for broadcasting advertisements. These 3 channels are primarily used by beacons and are chosen deliberately to minimize any collision with the WiFi channels. Beacon broadcasts its advertisement packet repetitively based on the selected advertising interval while hopping over the 3 designated channels \cite{vlugt}.

\subsection{Beacon Advantages for micro-location}
Beacons have several advantages to be used for micro-location.

\begin{itemize}
\item \textbf{Size.} Beacons are small in size hence, they can be placed almost in any indoor environment with no problem. They can be placed behind the ceiling, under objects or even on the walls.

\item \textbf{Energy efficiency.}  The great advantage of beacons comes from the energy efficient BLE protocol. At the same time, as the market of the beacons increases so does the different design approaches. There are small beacons that work with one single coin cell battery, there are beacons with two AA batteries  and there are  solar powered beacons \cite{spachos1}. The lifetime of these beacons can be up to two years without the need of battery replacement \cite{aislelabs}.

\item \textbf{Cost.} Most of the beacons in the market are cheap. Many beacons can be placed in a complex indoor environment to improve micro-location with  minimum cost.

\item \textbf{Inferences.} Beacons use BLE and they will not interfere  with other wireless infrastructures in the area.

\item \textbf{Passive mode.}  Beacons are broadcasters that do not do anything else besides sending a piece of information. The logic behind each signal is done by the supporting device, such as smartphones. Beacon signals are used by applications to trigger events and call actions, allowing  the users to interact with physical things. All the implementation is done on the device while the beacons just broadcast the signal.

\item \textbf{Platform independent.} Beacons can be used with iOS and Android devices.  Each platform required different protocols which have different packet layouts  but most platforms are able the listen to the different protocols.

\end{itemize}

\section{Using BLE Beacons for micro-location.}
In this section, a scenario where beacons can be used for micro-location is described, followed by some experimental results.
 
 \subsection{Test case.}
\begin{figure}[t!]
\centering%
\includegraphics[width=0.7\columnwidth]{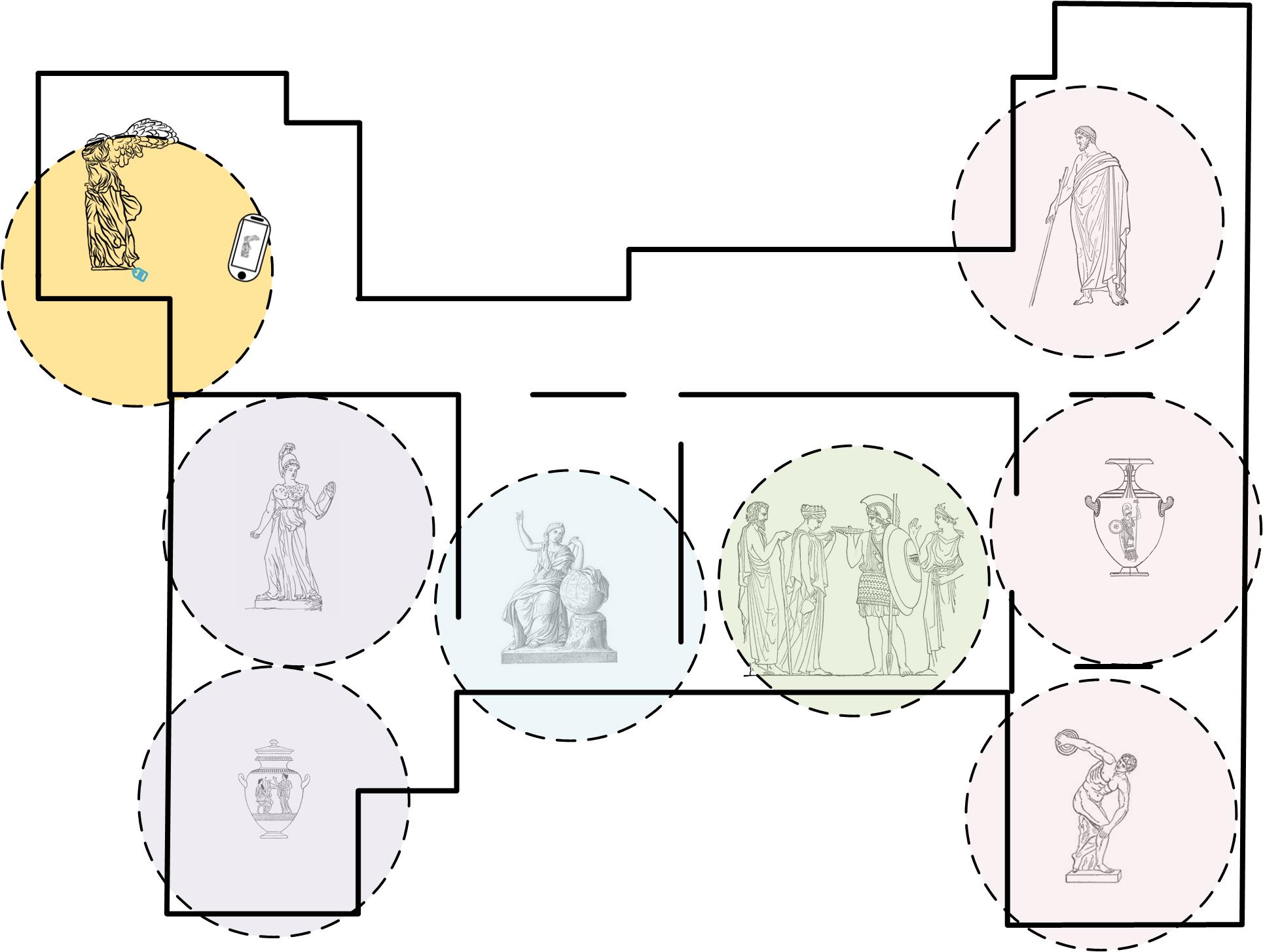}%
\caption{BLE beacons used in an interactive museum scenario.}%
\label{museum}%
\end{figure}

Museums and art galleries usually provide visitors either with paper booklets or with audio guides.  Unfortunately, interest may vary from person to person, while each visitors experience is also related to the available time to visit most of the exhibits.  Interactive and personalized museum tour needs to be developed. BLE beacons as a newly emerged technology can enhance visitors experience through micro-location, as shown in Fig. \ref{museum}.

Beacons can offer museums an opportunity to provide context to visitors through a smartphone application. Micro-location technology can make locating an exhibit much easier while at the same time  it can provide personalized  suggestions to the user, regarding the available exhibits. A mobile application can be developed which interacts with the available beacons. 

When visitors are close to an exhibit, they can get all the necessary information about the exhibit on their smartphone or BLE enabled mobile device in general. The application can also provide a recommendation to the visitor on the next exhibit they can visit, based on their current location and interest. At the same time, the application can provide an optimal tour of the museum based on each individual preferences. Beacon will also provide useful analytics to the museum. The number of visitors per exhibit can be collected, without violating visitors privacy. This analytics can be used to improve exhibits visibility.

The use of beacons provides several advantages for the museum and the visitors:
\begin{itemize}
\item \textbf{Promote exploration.}  The application can encourage users to visit exhibits in different places of the museum. Usually, visitors tend to spend most of their time in exhibits near to the entrance, missing the opportunity to explore exhibits across all the museum. Micro-location can help them identify the room they are interested in faster.
\item  \textbf{Personalized tour.} When a user is interested in an exhibit, the application can provide a guided tour based on that interest. An interactive and personalized tour with exhibits from the same chronological period or within the same interest category can be provided to the user, who would miss them without the application.  
\item  \textbf{Tour optimization.} For many visitors, the available time to spend in the museum is limited. The real-time analytics from the beacons can be used to provide an optimal route for the visitor, based on the available time for the visit.
\item  \textbf{Data analytics.} Beacons analytics can be used to improve general visitor experience. There are exhibits that are missed due to their location while there are exhibits that are overcrowded during a specific time of the day. Analytics can be used to optimize both cases and enhance visitors experience.
\end{itemize}

 \subsection{Experimental results.}
 In this subsection, we showcase the performance of the BLE beacons through a simple experimentation. We used BLE beacons from Gimbal Series 21 to examine the proximity estimation performance along with a smartphone which was used to collect the signals~\cite{mackey1}. The Kalman filter (KF-ST) was applied on the collected data offline. 

\begin{figure}[t!]
\centering%
\includegraphics[width=0.6\columnwidth]{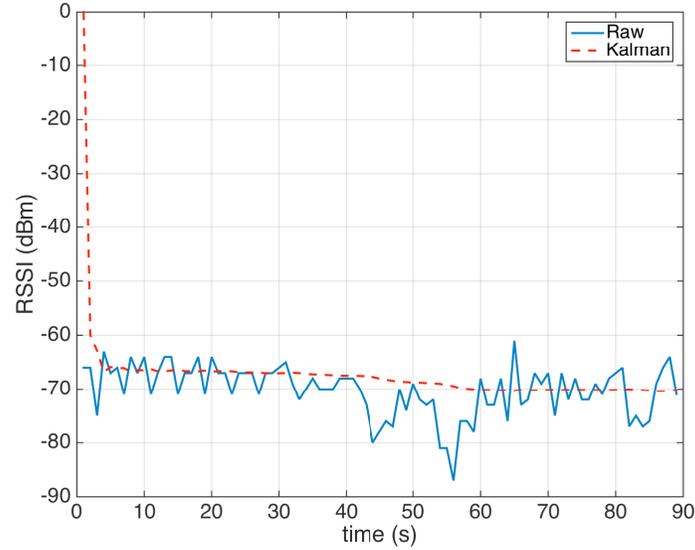}%
\caption{Received RSSI values  in two meters from the BLE beacon.}%
\label{kalman}%
\end{figure}

The Kalman filter estimation is shown in Fig.~\ref{kalman}. These are the collected RSSI values when the smartphone is 2 m away from the beacon. It is clear that Kalman filter can minimize the effect of interference between the beacon and the smartphone, such as when people are moving between the two communicating devices.

\begin{figure}[t!]
\centering%
\includegraphics[width=0.6\columnwidth]{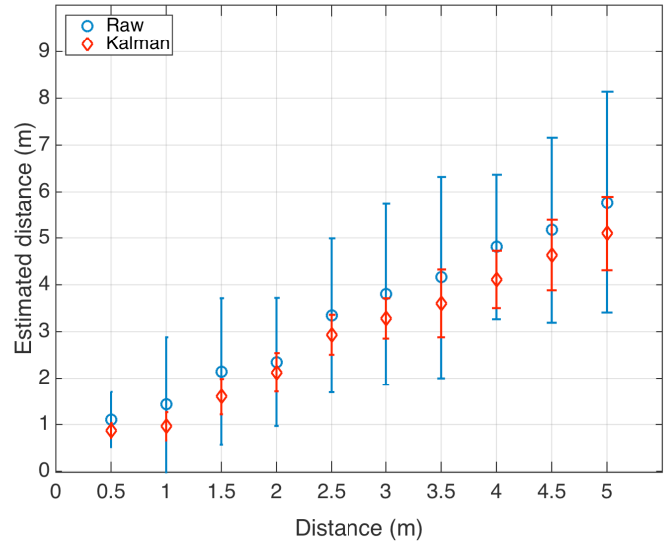}%
\caption{Distance estimation in ten different spots from the BLE beacon.}%
\label{RSSI}%
\end{figure}

To examine the performance of the Kalman filter, we placed the smartphone in ten different distances, starting from 50 cm and up to 5 m,  increasing the distance 50 cm every time. In every location, we collected data in the smartphone for approximately two minutes. The average RSSI values are shown in Fig.~\ref{RSSI}. When the smartphone is close to the beacon, the accuracy is high enough without filtering. As the distance increases, the accuracy without filtering decreases while the standard deviation of the data increases as well. Interference and noise affect the data transmission hence, as the distance between the communicating devices increase, these factors increase as well. Kalman filtering helps to keep the data close to the real value while the standard deviation is smaller. The use of Kalman filter helps minimize  the effect of random noise and interference during the experiment.

\begin{figure*}[t!]
\centering
\captionsetup[subfloat]{farskip=0pt}%
\subfloat[Raw data.]{\includegraphics[scale=0.35]{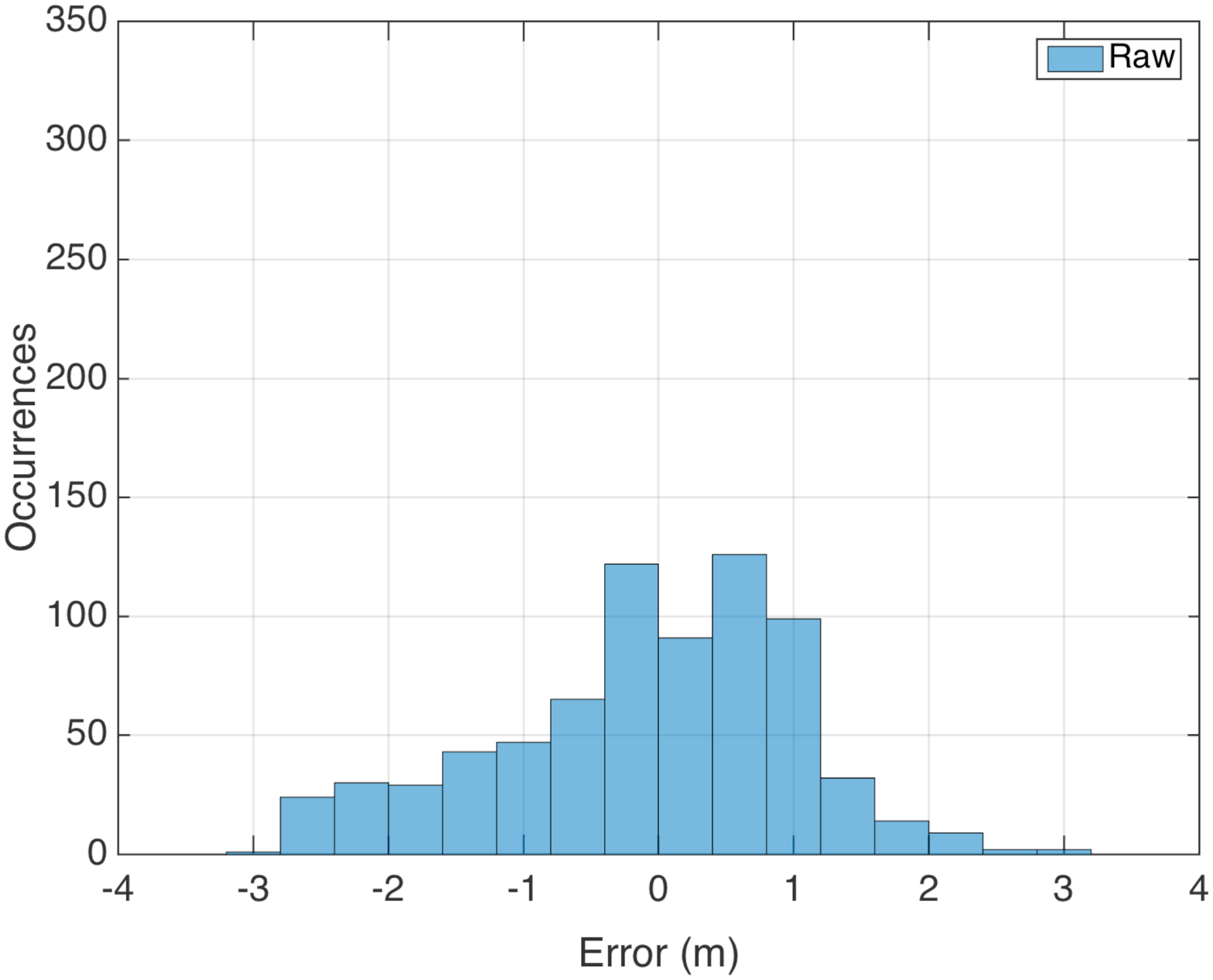}
\label{proximity}}
\subfloat[Kalman filter.]{\includegraphics[scale=0.35]{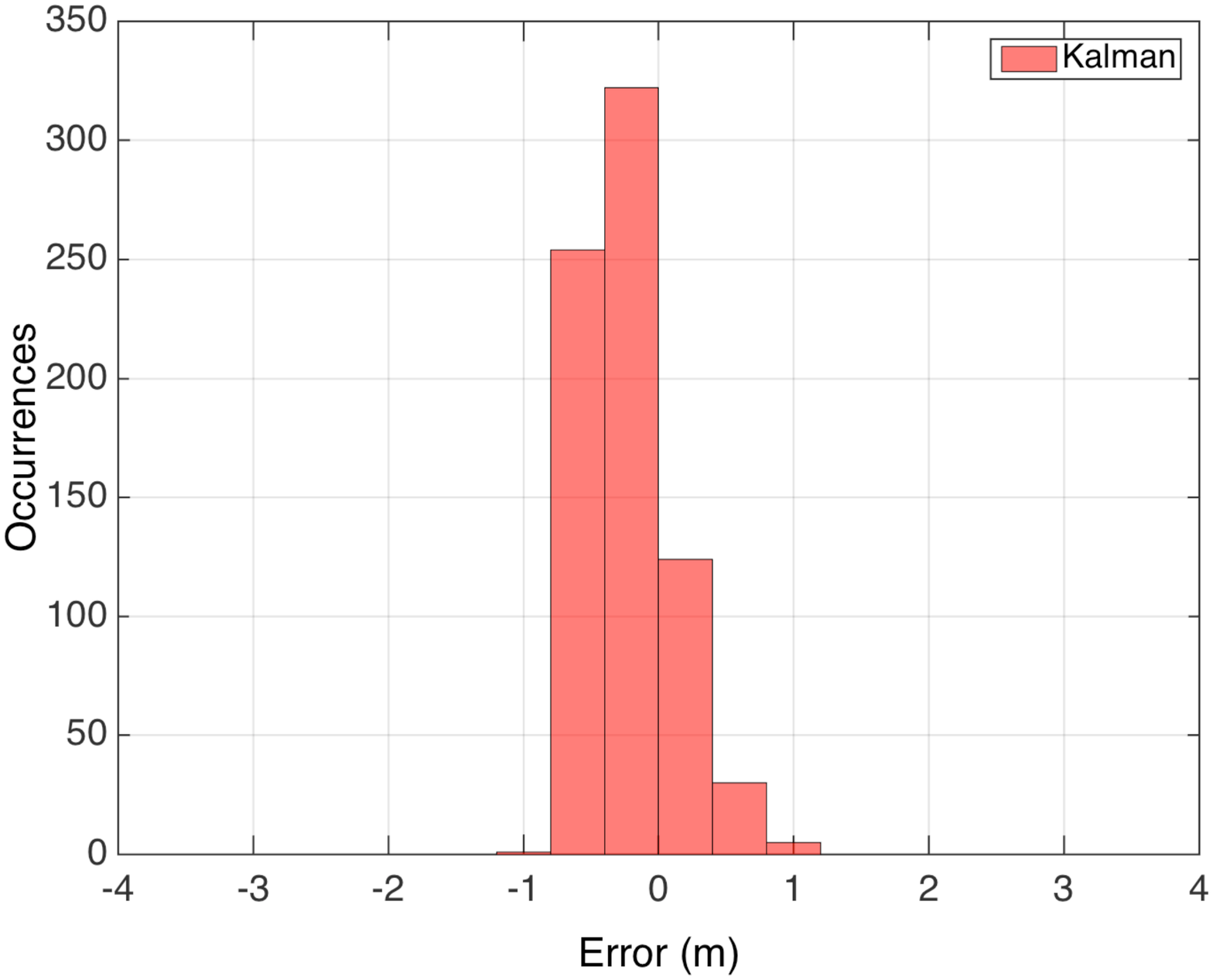}
\label{lateration}}
\caption{Histogram of experimental error.}
\label{estimation}
\end{figure*}

We further examine  the error between the estimated distance and the real distance, and the number of occurrences of each group of errors  as shown in Fig.~\ref{estimation}. Without filtering, the error is within 3 m from the real location when distances up to 5 m are tested. In many  applications that use micro-location such as the test case, the location error should be smaller. A smaller error comes when Kalman filter is used. The error is within 1 m from the real location, which can be acceptable from many micro-location applications.

\section{Concluding Remarks.}

This paper provides an overview of  wireless technologies that can be used for micro-location in smart buildings, with the use of IoT devices. BLE is among the most energy-efficient technologies. BLE beacons are small, low-cost devices that can be used for localization. Unfortunately, they are prone to interference due to their wireless nature. Signal processing techniques, such as Kalman filters can be used to enhance their performance. 

A case study of BLE beacons in an interactive museum is also discussed. According to the experimental results, signal processing techniques can enhance beacons performance and provide accurate micro-location in the era of IoT.
 
\section*{Author Biographies}

\noindent
\textbf{Petros Spachos} (\textit{IEEE Senior Member, Ph.D., University of Toronto, 2014})  is an Assistant Professor at School of Engineering, University of Guelph, Canada. He received the Diploma degree in Electronic and Computer Engineering from the Technical University of Crete, Greece, and the M.A.Sc. and the Ph.D. degree  both in Electrical and Computer Engineering from the University of Toronto, Canada. His research interests include experimental
wireless networking and mobile computing
with a focus on wireless sensor networks, smart cities and the Internet of
Things. He is a senior member of the IEEE.

\noindent
\textbf{Ioannis Papapanagiotou} (\textit{IEEE Senior Member, Ph.D., NC State University, 2012}) is an engineering manager at Netflix, a research assistant professor at the University of New Mexico, a graduate faculty at Purdue University, and a mentor at the International Accelerator. He holds a dual Ph.D. degree in Computer Engineering and Operations Research. His main focus is on distributed systems, cloud computing, and the Internet of Things. He has been awarded the NetApp faculty fellowship and established an Nvidia CUDA Research Center at Purdue University. Ioannis has also received the IBM Ph.D. Fellowship and Academy of Athens Ph.D. Fellowship for his Ph.D. research, and best paper awards in several IEEE conferences for his academic contributions. Ioannis has authored a number of research articles and patents. Ioannis is a senior member of ACM and IEEE.

\noindent
\textbf{Konstantinos N. Plataniotis} (\textit{IEEE Fellow, Ph.D.,  Florida Institute of Technology, 1997})  is a Professor and Bell Canada Chair in Multimedia with the ECE Department at the University of Toronto. He is a registered professional engineer in Ontario, Fellow of the IEEE and Fellow of the Engineering Institute of Canada. He has served as Signal Processing Society Vice President for Membership (2014 -2016) and the Editor-in-Chief for IEEE Signal Processing Letters (2009-2011). He was Technical Co-Chair for ICASSP 2013 and the General Co-Chair for
2017 IEEE GlobalSIP.  He co-Chairs the 2018 IEEE International Conference
on Image Processing, and the 2021 IEEE International Conference on
Acoustics, Speech and Signal Processing.

    \vspace{-2ex}

\bibliographystyle{IEEEtran}
\bibliography{IEEEabrv,SPMbib}

\begin{thebibliography}{10}
\providecommand{\url}[1]{#1}
\csname url@samestyle\endcsname
\providecommand{\newblock}{\relax}
\providecommand{\bibinfo}[2]{#2}
\providecommand{\BIBentrySTDinterwordspacing}{\spaceskip=0pt\relax}
\providecommand{\BIBentryALTinterwordstretchfactor}{4}
\providecommand{\BIBentryALTinterwordspacing}{\spaceskip=\fontdimen2\font plus
\BIBentryALTinterwordstretchfactor\fontdimen3\font minus
  \fontdimen4\font\relax}
\providecommand{\BIBforeignlanguage}[2]{{%
\expandafter\ifx\csname l@#1\endcsname\relax
\typeout{** WARNING: IEEEtran.bst: No hyphenation pattern has been}%
\typeout{** loaded for the language `#1'. Using the pattern for}%
\typeout{** the default language instead.}%
\else
\language=\csname l@#1\endcsname
\fi
#2}}
\providecommand{\BIBdecl}{\relax}
\BIBdecl

\bibitem{sadowski}
S.~Sadowski and P.~Spachos, ``Rssi-based indoor localization with the internet
  of things,'' \emph{IEEE Access}, pp. 1--1, 2018.

\bibitem{smartBuildings}
F.~Zafari, I.~Papapanagiotou, and K.~Christidis, ``Microlocation for
  internet-of-things-equipped smart buildings,'' \emph{IEEE Internet of Things
  Journal}, vol.~3, no.~1, pp. 96--112, Feb. 2016.

\bibitem{SikeMicro}
D.~Sikeridis, B.~P. Rimal, I.~Papapanagiotou, and M.~Devetsikiotis,
  ``Unsupervised crowd-assisted learning enabling location-aware facilities,''
  \emph{IEEE Internet of Things Journal}, pp. 1--1, 2018.

\bibitem{ibeaconpacket}
\BIBentryALTinterwordspacing
Apple. (2014, June 2) Getting started with ibeacon. [Online]. Available:
  \url{https://developer.apple.com/ibeacon/Getting-Started-with-iBeacon.pdf}
\BIBentrySTDinterwordspacing

\bibitem{eddystonepacket}
\BIBentryALTinterwordspacing
Google. (2017, July 5) Google eddystone format. [Online]. Available:
  \url{https://developers.google.com/beacons/eddystone}
\BIBentrySTDinterwordspacing

\bibitem{bordoy}
J.~Bordoy, A.~Traub-Ens, A.~Sadr, J.~Wendeberg, F.~Höflinger, C.~Schindelhauer,
  and L.~Reindl, ``Bank of kalman filters in closed-loop for robust
  localization using unsynchronized beacons,'' \emph{IEEE Sensors Journal},
  vol.~16, no.~19, pp. 7142--7149, Oct. 2016.

\bibitem{zafari2017ibeacon}
F.~Zafari, I.~Papapanagiotou, M.~Devetsikiotis, and T.~Hacker, ``An ibeacon
  based proximity and indoor localization system,'' \emph{arXiv preprint
  arXiv:1703.07876v2}, 2017.

\bibitem{particleFilt}
F.~Zafari and I.~Papapanagiotou, ``Enhancing ibeacon based micro-location with
  particle filtering,'' in \emph{2015 IEEE Global Communications Conference
  (GLOBECOM)}, Dec. 2015, pp. 1--7.

\bibitem{NN_knB}
C.~Takahashi and K.~Kondo, ``Accuracy evaluation of an indoor positioning
  method using ibeacons,'' in \emph{2016 IEEE 5th Global Conference on Consumer
  Electronics}, Oct. 2016, pp. 1--2.

\bibitem{PDR}
Z.~Chen, Q.~Zhu, H.~Jiang, and Y.~C. Soh, ``Indoor localization using
  smartphone sensors and ibeacons,'' in \emph{2015 IEEE 10th Conference on
  Industrial Electronics and Applications (ICIEA)}, June 2015, pp. 1723--1728.

\bibitem{IBE}
\BIBentryALTinterwordspacing
{Institute for Building Efficiency}. [Online]. Available:
  \url{http://www.buildingefficiencyinitiative.org/}
\BIBentrySTDinterwordspacing

\bibitem{wifi}
\BIBentryALTinterwordspacing
{The Working Group for WLAN Standards}. {IEEE 802.11 Wireless Local Area
  Networks}. [Online]. Available: \url{http://www.ieee802.org/11/}
\BIBentrySTDinterwordspacing

\bibitem{zigbee}
\BIBentryALTinterwordspacing
P.~Baronti, P.~Pillai, V.~W. Chook, S.~Chessa, A.~Gotta, and Y.~F. Hu,
  ``Wireless sensor networks: A survey on the state of the art and the 802.15.4
  and zigbee standards,'' \emph{Computer Communications}, vol.~30, no.~7, pp.
  1655 -- 1695, 2007, wired/Wireless Internet Communications. [Online].
  Available:
  \url{http://www.sciencedirect.com/science/article/pii/S0140366406004749}
\BIBentrySTDinterwordspacing

\bibitem{rfid}
R.~Want, ``An introduction to rfid technology,'' \emph{IEEE Pervasive
  Computing}, vol.~5, no.~1, pp. 25--33, Jan. 2006.

\bibitem{blue5}
\BIBentryALTinterwordspacing
{Bluetooth Special Interest Group}. Bluetooth 5.0 core specification. [Online].
  Available:
  \url{https://www.bluetooth.com/specifications/bluetooth-core-specification/bluetooth5}
\BIBentrySTDinterwordspacing

\bibitem{kusens2017electronic}
M.~Kusens, ``Electronic location identification and tracking system with beacon
  clustering,'' Sep.~26 2017, uS Patent 9,774,991.

\bibitem{halow}
\BIBentryALTinterwordspacing
E.~Khorov, A.~Lyakhov, A.~Krotov, and A.~Guschin, ``A survey on ieee 802.11ah:
  An enabling networking technology for smart cities,'' \emph{Computer
  Communications}, vol.~58, pp. 53 -- 69, 2015, special Issue on Networking and
  Communications for Smart Cities. [Online]. Available:
  \url{http://www.sciencedirect.com/science/article/pii/S0140366414002989}
\BIBentrySTDinterwordspacing

\bibitem{lorawan}
\BIBentryALTinterwordspacing
{Lora Alliance Technology}. Lorawan. [Online]. Available:
  \url{https://lora-alliance.org/about-lorawan}
\BIBentrySTDinterwordspacing

\bibitem{kennedy}
B.~Kennedy, G.~Taylor, and P.~Spachos, ``Ble beacon based patient tracking in
  smart care facilities,'' in \emph{2018 IEEE International Conference on
  Pervasive Computing and Communications Workshops (PerCom Workshops)}, March
  2018.

\bibitem{Lifi}
M.~Ayyash, H.~Elgala, A.~Khreishah, V.~Jungnickel, T.~Little, S.~Shao,
  M.~Rahaim, D.~Schulz, J.~Hilt, and R.~Freund, ``Coexistence of wifi and lifi
  toward 5g: concepts, opportunities, and challenges,'' \emph{IEEE
  Communications Magazine}, vol.~54, no.~2, pp. 64--71, February 2016.

\bibitem{arulampalam2002tutorial}
M.~S. Arulampalam, S.~Maskell, N.~Gordon, and T.~Clapp, ``A tutorial on
  particle filters for online nonlinear/non-gaussian bayesian tracking,''
  \emph{Signal Processing, IEEE Transactions on}, vol.~50, no.~2, pp. 174--188,
  2002.

\bibitem{djuric2003particle}
P.~M. Djuric, J.~H. Kotecha, J.~Zhang, Y.~Huang, T.~Ghirmai, M.~F. Bugallo, and
  J.~Miguez, ``Particle filtering,'' \emph{Signal Processing Magazine, IEEE},
  vol.~20, no.~5, pp. 19--38, 2003.

\bibitem{guveenhancements}
I.~Guvenc, C.~Abdallah, R.~Jordan, and O.~Dedoglu, ``Enhancements to {RSS}
  based indoor tracking systems using kalman filters,'' in \emph{In GSPx \&
  International Signal Processing Conference}, 2003.

\bibitem{kriz}
F.~M. P.~Kriz and T.~Kozel, ``Improving indoor localization using bluetooth low
  energy beacons,'' in \emph{Mobile Information Systems}, March 2016, pp. 1--7.

\bibitem{altbeacons}
\BIBentryALTinterwordspacing
AltBeacon. (2016, June 7) Specifications on altbeacon. [Online]. Available:
  \url{https://github.com/AltBeacon/spec}
\BIBentrySTDinterwordspacing

\bibitem{geobeacon}
\BIBentryALTinterwordspacing
Geobeacon. [Online]. Available: \url{https://github.com/Tecno-World/GeoBeacon}
\BIBentrySTDinterwordspacing

\bibitem{aislelabs}
\BIBentryALTinterwordspacing
Aislelabs. (2015) The hitchhikers guide to ibeacon hardware: A comprehensive
  report. [Online]. Available: \url{http://www.aislelabs.com/
  reports/beacon-guide/}
\BIBentrySTDinterwordspacing

\bibitem{vlugt}
\BIBentryALTinterwordspacing
E.~Vlugt. (2013) Bluetooth low energy, beacons and retail. [Online]. Available:
  \url{, http://
  global.verifone.com/media/3603729/bluetooth-low-energy-beacons-retail-wp.pdf}
\BIBentrySTDinterwordspacing

\bibitem{spachos1}
\BIBentryALTinterwordspacing
P.~Spachos and A.~Mackey, ``Energy efficiency and accuracy of solar powered ble
  beacons,'' \emph{Computer Communications}, vol. 119, pp. 94 -- 100, 2018.
  [Online]. Available:
  \url{http://www.sciencedirect.com/science/article/pii/S0140366417309891}
\BIBentrySTDinterwordspacing

\bibitem{mackey1}
A.~Mackey and P.~Spachos, ``Performance evaluation of beacons for indoor
  localization in smart buildings,'' in \emph{2017 IEEE Global Conference on
  Signal and Information Processing (GlobalSIP)}, Nov. 2017, pp. 823--827.

\end{thebibliography}

\end{document}